\documentclass[preprint,preprintnumbers,12pt]{elsarticle}

\usepackage{amssymb}
\usepackage{amsmath}
\usepackage{graphicx}
\usepackage{dcolumn}
\usepackage{bm}
\usepackage{braket}
\usepackage{hyperref}

\graphicspath{{./Figures/}}

\newcommand{\be}{\begin{equation}}
\newcommand{\ee}{\end{equation}}
\newcommand{\fn}{\footnote}
\newcommand{\LQ}{\Lambda_{\rm QCD}}

\newcommand{\lt}{\left}
\newcommand{\rt}{\right}
\newcommand{\bz}{\beta_0}
\newcommand{\non}{\nonumber \\}

\newcommand{\IR}{\rm IR}
\newcommand{\UV}{\rm UV}
\newcommand{\US}{\rm US}

\newcommand{\DV}{\Delta V}

\journal{Physics Letter B}

\begin{document}


\begin{frontmatter}



\title{Renormalon free part of an ultrasoft correction \\
to the static QCD potential 
}


\author{Hiromasa Takaura}

\address{Department of Physics, Tohoku University,
Sendai, 980--8578 Japan
}

\begin{abstract}
Perturbative calculations of the static QCD potential
have the $u=3/2$ renormalon uncertainty.
In the multipole expansion performed within pNRQCD,
this uncertainty at LO is known to get canceled against 
the ultrasoft correction at NLO.
To investigate the net contribution remaining after this renormalon cancellation,
we propose a formulation to separate the ultrasoft correction
into renormalon uncertainties and a renormalon independent part.
We focus on very short distances $\LQ r \lesssim 0.1$
and investigate the ultrasoft correction 
based on its perturbative evaluation in the large-$\bz$ approximation. 
We also propose a method to examine the local gluon condensate,
which appears as the first nonperturbative effect to the static QCD potential,
without suffering from the $u=2$ renormalon.
\end{abstract}


\end{frontmatter}

\section{Introduction}
\noindent
The static QCD potential plays an important role to investigate 
the QCD dynamics.
It has been investigated extensively by using perturbation theory,
effective field theory and lattice simulations. 

In perturbative evaluations, 
perturbative coefficients are expected to show factorial behaviors at large orders.
Such divergent behaviors, in particular those related to a positive renormalon, 
induce ambiguity to the resummation of the perturbative series \cite{Beneke:1998ui}.  
For the static QCD potential, renormalons are located at positive half integers in the Borel $u$-plane. 
The first renormalon at $u=1/2$ causes an uncertainty to the $r$-independent constant of the potential.
This renormalon is known to get canceled in the total energy 
(i.e. the sum of the QCD potential and twice the pole mass)  within usual perturbation theory
once the pole mass is expressed as a perturbative series in terms of the $\overline{\rm MS}$ mass.

In considering cancellation of the other renormalons,
it is useful to adopt the effective field theory (EFT)
known as potential non-relativistic QCD (pNRQCD) \cite{Brambilla:1999xf}.
The renormalons are expected to get canceled ultimately
in the multipole expansion performed in this EFT.
The leading order (LO) term of this expansion is the singlet potential $V_S(r)$,
which behaves as $\mathcal{O}(1/r)$ and can be evaluated in perturbation theory.
The renormalons for $V_S$ are located at positive half integers as mentioned above,
and the leading $r$-dependent uncertainty is caused by the $u=3/2$ renormalon. 
The next-to-leading order (NLO) term in the multipole expansion, which we denote by $\delta E_{\US}$,
represents the dynamics at the ultrasoft scale $\sim \alpha_s(r^{-1})/r$ and its explicit $r$-dependence is $\mathcal{O}(r^2)$.
In Ref.~\cite{Brambilla:1999xf}, it was pointed out within the leading-logarithmic approximation that 
the $u=3/2$ renormalon exists in $\delta E_{\US}$ and 
it cancels against the $u=3/2$ renormalon of $V_S$.
In Ref.~\cite{Sumino:2004ht}, the perturbative evaluation of $\delta E_{\US}$ in 
the large-$\bz$ approximation was completed, and again, 
the $u=3/2$ renormalon cancellation was confirmed.

Although the $u=3/2$ renormalon cancellation has been established,
it has not been clarified what remains in the NLO calculation, $V_S+\delta E_{\US}$, 
as a consequence of this renormalon cancellation.
In particular, the net contribution to $\delta E_{\US}$ remaining after this cancellation has not been made clear.
In this Letter, we investigate the net contribution to $\delta E_{\US}$,
which is not affected by the renormalon cancellation.
The distances considered here are $\LQ r \lesssim 0.1$,
where $\delta E_{\US}$ as well as $V_S$ can be evaluated perturbatively since the ultrasoft scale satisfies $\alpha_s(r^{-1})/r \gg \LQ$.
In this range, the leading nonperturbative correction is given through the local gluon condensate.\fn{
The appearance of this nonperturbative effect has been first considered in Refs.~\cite{Voloshin:1978hc, Leutwyler:1980tn,Flory:1982qx}, 
and can be understood in a systematic expansion of pNRQCD \cite{Brambilla:1999xf}.}
In order to examine the local gluon condensate, the perturbative part, i.e. $V_S+\delta E_{\US}$, should be clearly known in advance.
Although the currently available order of perturbative expansion is 
far from the (expected) relevant order to the $u=3/2$ renormalon,\fn{
For the singlet potential $V_S$, the relevant perturbation order to the $u=3/2$ renormalon is roughly estimated as $n_*=\frac{6 \pi}{\bz \alpha_s(r^{-1})}\sim 20$, while the exact series is currently known up to $\mathcal{O}(\alpha_s^4)$.}
this Letter aims at promoting theoretical understanding of the static QCD potential without suffering from renormalons.

To investigate the net NLO correction, we propose a formulation to separate $\delta E_{\US}$
into its renormalon uncertainties and a renormalon free part.
In perturbative evaluations, the large-$\bz$ approximation is used.
For the LO term $V_S$, a renormalon separation has been performed in Ref.~\cite{Sumino:2005cq}.
In this Letter, we will see that the $u=3/2$ renormalon uncertainty of $V_S$ separated out in Ref.~\cite{Sumino:2005cq}
is canceled against that of $\delta E_{\US}$ identified here.
As a result, a theoretical expression after the $u=3/2$ renormalon cancellation is obtained
by the sum of the renormalon free parts of $V_S$ and $\delta E_{\US}$.\fn{
The $u=1/2$ renormalon uncertainty in $V_S$ is just omitted 
as it changes only the $r$-independent constant.}
Each is presented in an analytic form in this Letter.\fn{
The analytical result for $V_S$ that is free from renormalons has been given 
in Ref.~\cite{Sumino:2005cq}.}

Once the renormalon separation of $\delta E_{\US}$ is performed,
it is straightforward to cope with the residual renormalon at $u=2$.
This renormalon in $\delta E_{\US}$ (found explicitly in Ref.~\cite{Sumino:2004ht})
is consistent with the fact that the local gluon condensate appears as the first nonperturbative effect.
The $u=2$ renormalon induces an $\mathcal{O}(\LQ^4 r^3)$ error to $V_S+\delta E_{\US}$,
which is the same magnitude as the term of the local gluon condensate.
Hence, the $u=2$ renormalon is an obstacle in examining the local gluon condensate even after the $u=3/2$ renormalon is removed. 
We circumvent this problem by including the $u=2$ renormalon uncertainty of $\delta E_{\US}$ in the local gluon condensate,
which results in the cancellation of the $u=2$ renormalon in the local gluon condensate.
This renormalon cancellation is explicitly confirmed in this Letter using the large-$\bz$ approximation.
As a result, one can obtain the expansion in $r$ up to the order including the local gluon condensate
such that each term does not have the $u=3/2$ and $u=2$ renormalons.
Such a result can be used to extract the local gluon condensate numerically, for instance, 
by comparing lattice simulations with the calculation presented here.

Our formulation to extract a renormalon free part of $\delta E_{\US}$  
is an extension of Refs.~\cite{Mishima:2016xuj, Mishima:2016vna},
which propose the method to extract a renormalon free part
in the leading term in operator product expansion (see Ref.~\cite{Ball:1995ni} as a related work).
The characteristics of our formulation is to introduce explicit cutoff scales,
which are compatible with the concept of the EFT.\fn{
The cutoff scale dependence vanishes in the final results.}
This clarifies intuitively how renormalon uncertainties appear
and also vanish when combined with contributions of different energy scales. 
In particular, we will see how a renormalon free part is identified 
in connection with the cutoff scales.

\section{Extraction of renormalon free part}
\begin{figure}[tbhp]
\begin{minipage}{0.5\hsize}
\begin{center}
\includegraphics[width=3.8cm,height=4cm,scale=1]{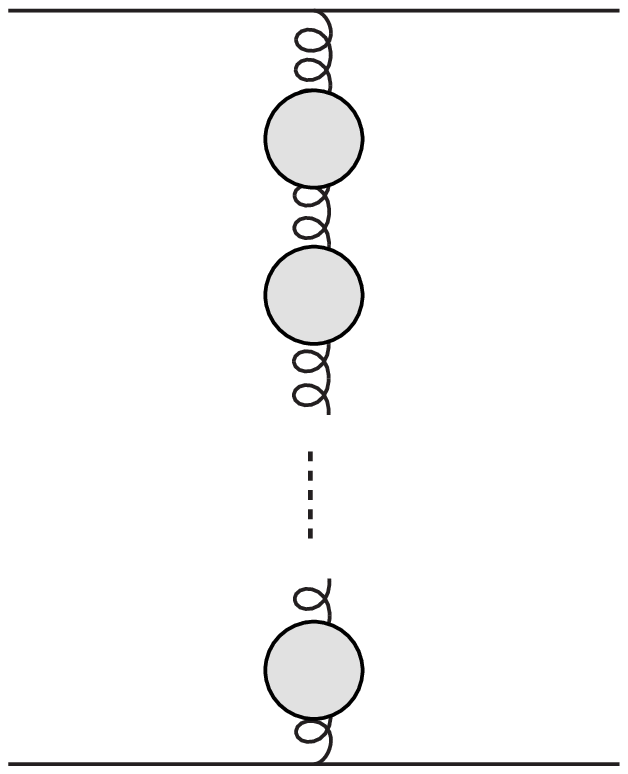}
\end{center}
\end{minipage}
\begin{minipage}{0.5\hsize}
\begin{center}
\includegraphics[width=7cm]{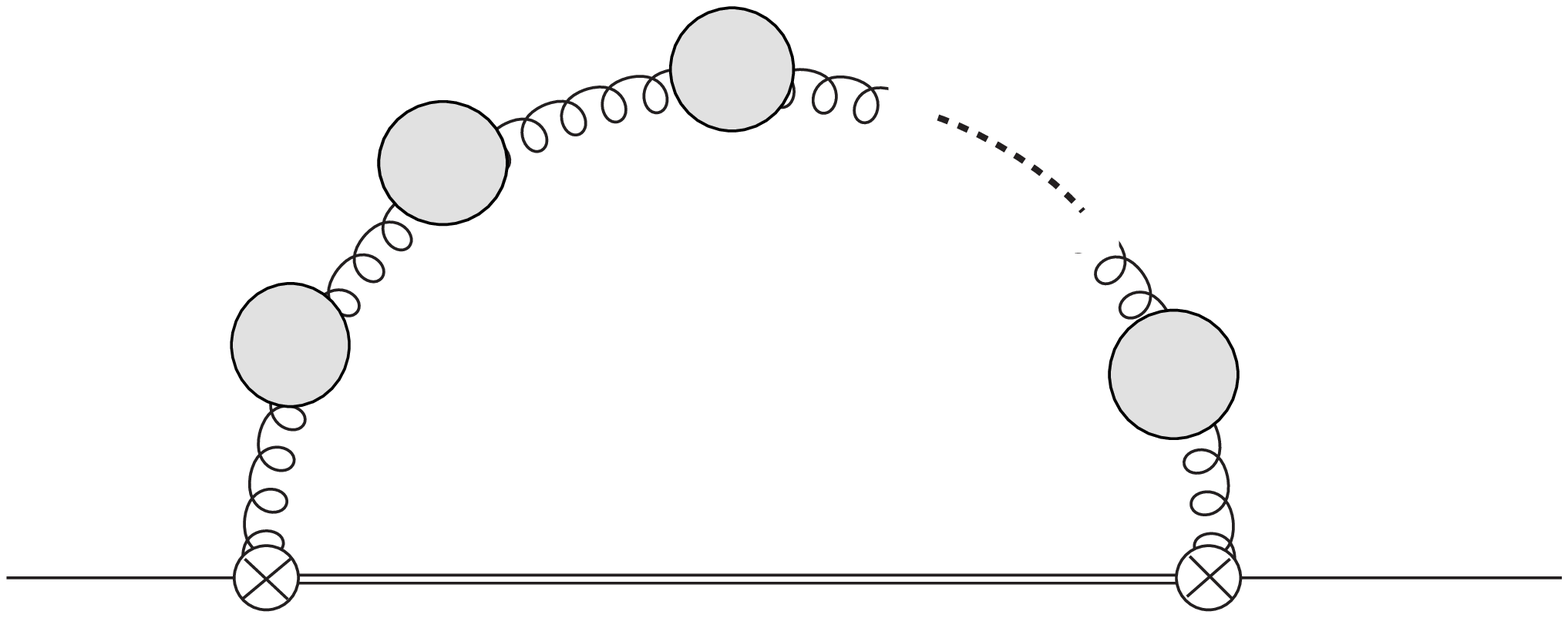}
\end{center}
\end{minipage}
\caption{Diagrams for the singlet potential (left) and $\delta E_{\US}$ (right) in the large-$\bz$ approximation.}
\label{fig:Diagrams}
\end{figure}

In the multipole expansion performed within pNRQCD, 
the static QCD potential is represented as \cite{Brambilla:1999xf}
\be
V_{\rm QCD}(r)=V_S(r) + \delta E_{\US}(r)+\dots \label{multipoleexp} \, ,
\ee
\be
\delta E_{\US} (r)=-i \frac{4 \pi \alpha_s}{N_c} T_F \int_0^{\infty} dt \, e^{-i \Delta V(r) t} 
\braket{\vec{r} \cdot \vec{E}^a(t) \varphi_{\rm adj}(t,0)^{ab} \vec{r} \cdot \vec{E}^b(0)} \, , \label{deltaEUS}
\ee
where the dots denote higher order corrections in $r$;
$\Delta V(r)=V_O(r)-V_S(r)$ denotes the difference between the octet and singlet potentials,
which specifies the ultrasoft scale;
$\vec{E}^a$ is the ultrasoft chromoelectric field.
See Ref.~\cite{Brambilla:1999xf} for details.
In the following, we evaluate $V_S$ and $\delta E_{\US}$ in 
perturbation theory especially using the large-$\bz$ approximation
\cite{Beneke:1994qe, Broadhurst:1993ru}.
The corresponding diagrams are shown in Fig.~\ref{fig:Diagrams}.

Let us first sketch what we will do in the following. 
We introduce cutoff scales $\mu_1$ and $\mu_2$
to divide the energy region: $\LQ \ll \mu_2 \ll \DV \ll \mu_1 \ll r^{-1}$. 
We define $V_S$ as a soft quantity
by restricting the gluon momentum to be higher than $\mu_1$.
Similarly, we define $\delta E_{\US}$ as an ultrasoft quantity
by requiring the relevant momentum $p$ to be $\mu_2 < p<\mu_1$.
Accordingly, we perform the multipole expansion as
\be
V_{\rm QCD}(r)=V_S(r;\mu_1)+\delta E_{\US}(r;\mu_1,\mu_2)+\dots \, . \label{multi2}
\ee
For the singlet potential $V_S(r;\mu_1)$, 
we follow the separation performed in Ref.~\cite{Sumino:2005cq}:\fn{
In Eq.~\eqref{VS}, we omit the $r$-independent constant related to the $u=1/2$ renormalon .
\label{fn:VSconst}}
\be
V_S(r;\mu_1)=V_S^{\rm RF}(r)+\mathcal{C}_2(\mu_1) r^2 +\mathcal{O}(r^3) \, , \label{VS}
\ee
where $V_S^{\rm RF}$ is a renormalon free part and 
has a Coulomb+linear form. 
The leading cutoff dependence in $V_S(r,\mu_1)$, $\mathcal{C}_2(\mu_1) r^2 \sim \mu_1^3 r^2$,
is caused by the $u=3/2$ renormalon.
In this Letter, we show that $\delta E_{\US}(r;\mu_1,\mu_2)$ can be decomposed as
\be
\delta E_{\US}(r;\mu_1,\mu_2)
\sim \delta E_{\US}^{\rm RF}(r)-\mathcal{C}_2(\mu_1) r^2
+\mathcal{O}(\mu_2^4 r^3) \, , \label{anti}
\ee
where $\delta E_{\US}^{\rm RF}(r)$ is  independent of $\mu_1$ and $\mu_2$
and is free from renormalons.
As a result, in the multipole expansion \eqref{multi2},
we have
\be
V_{\rm QCD}(r) = V_S^{\rm RF}(r)+\delta E_{\US}^{\rm RF}(r)+\mathcal{O}(r^3) \, .
\ee
In this way, we can obtain a net contribution up to NLO, where each term does not have renormalon ambiguity.

\begin{figure}[htbp]
\begin{minipage}{0.3\hsize}
\begin{center}
\includegraphics[width=4.5cm]{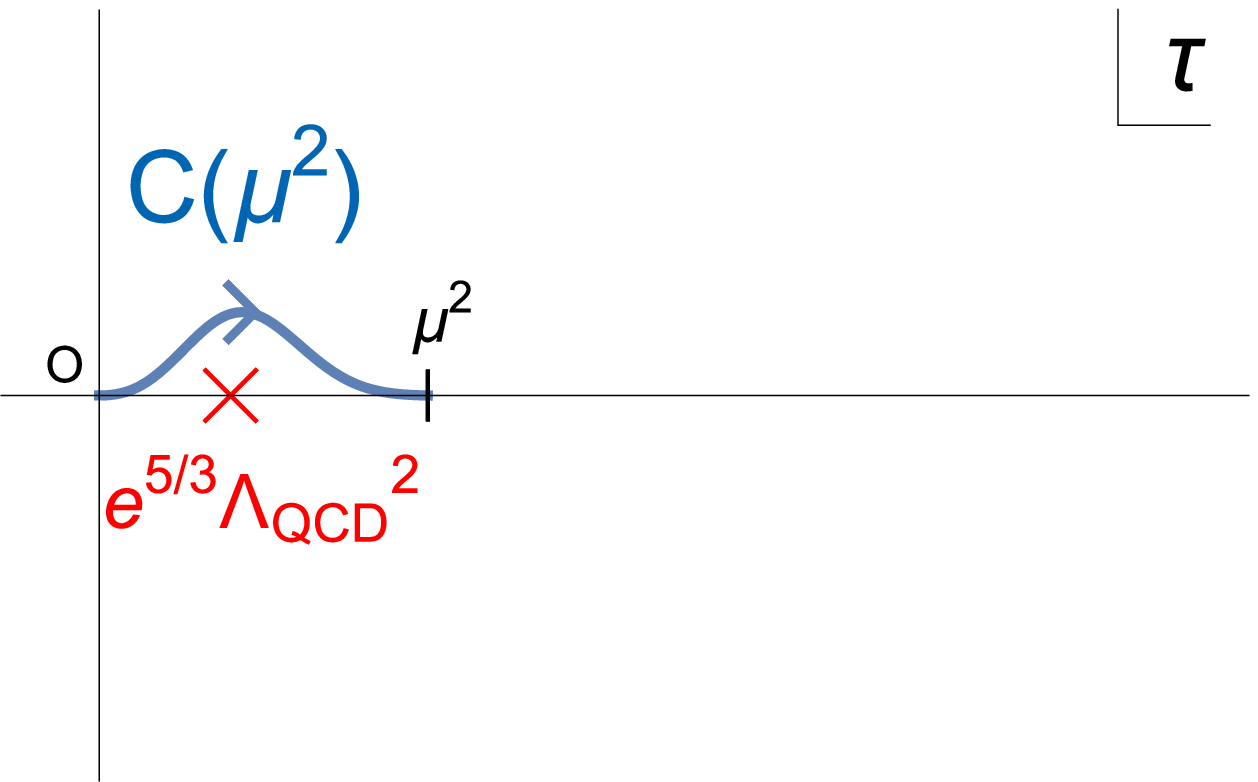}
\end{center}
\end{minipage}
\begin{minipage}{0.3\hsize}
\begin{center}
\includegraphics[width=4.5cm]{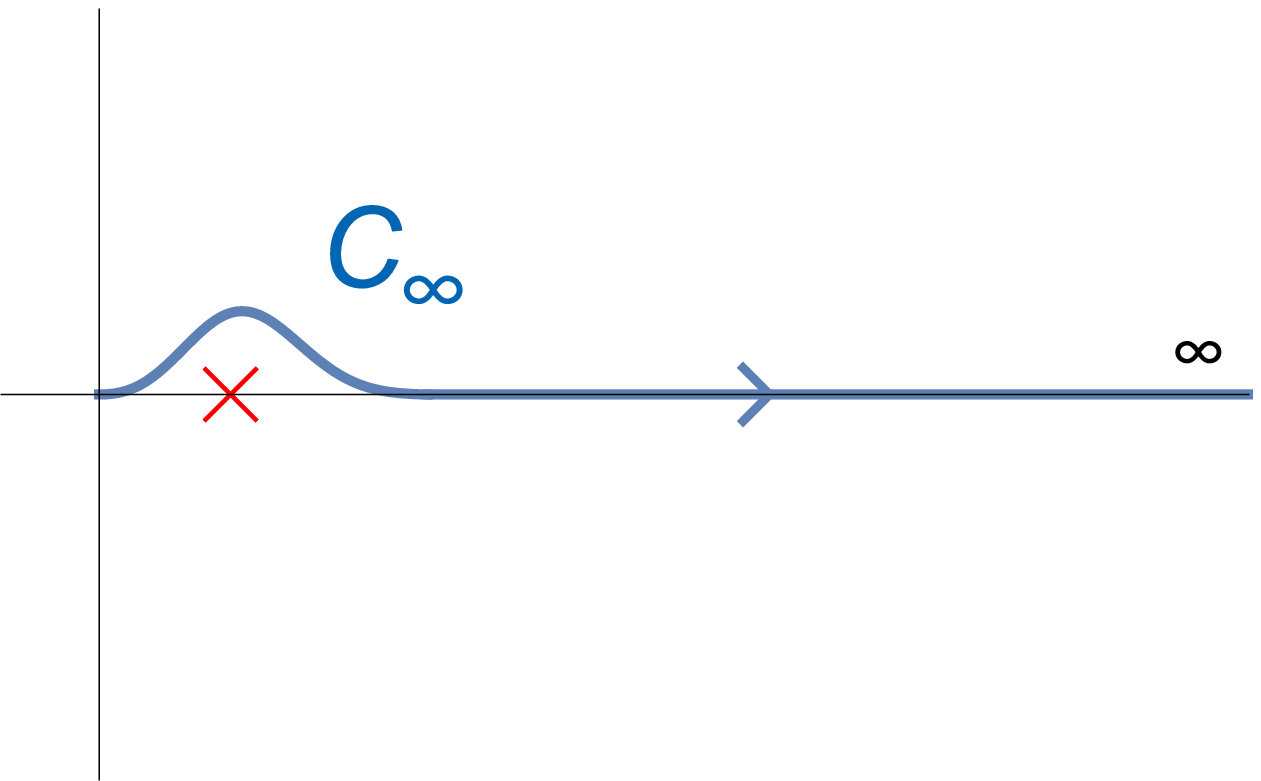}
\end{center}
\end{minipage}
\begin{minipage}{0.3\hsize}
\begin{center}
\includegraphics[width=4.5cm]{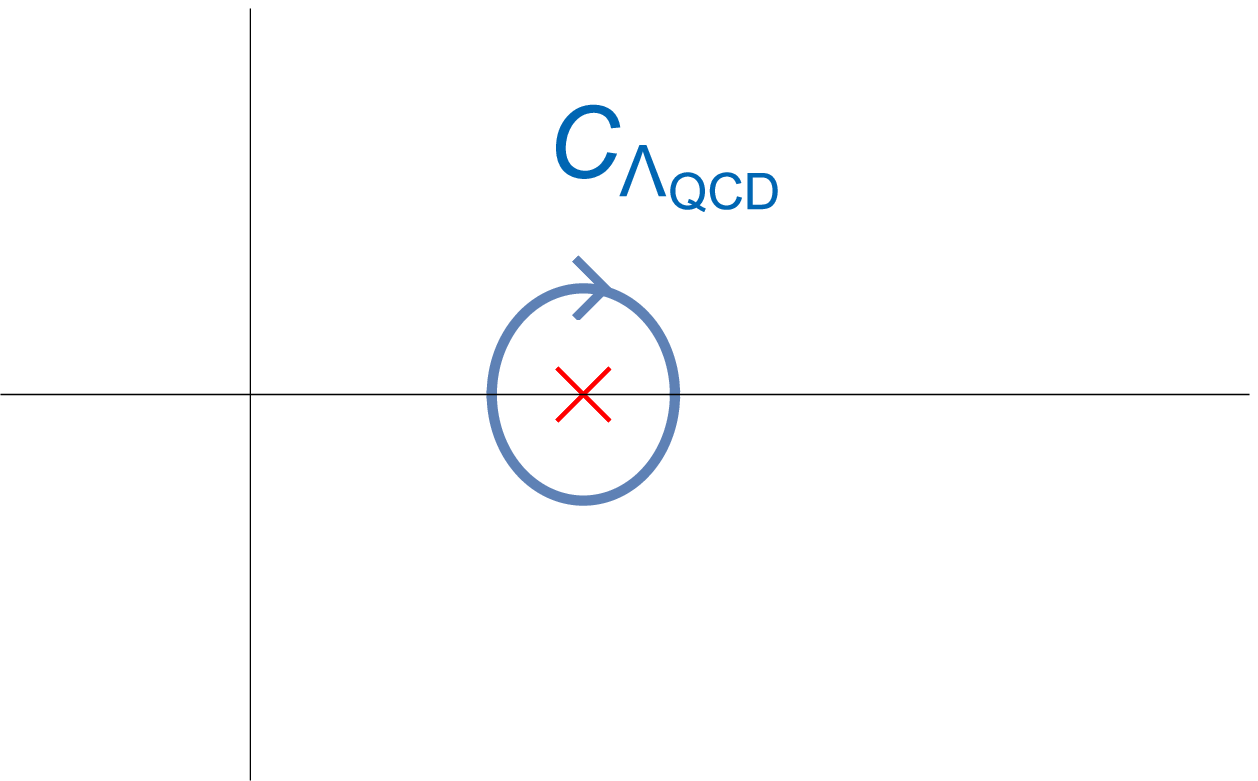}
\end{center}
\end{minipage}
\caption{Contours in complex $\tau$-plane.}
\label{fig:contours}
\end{figure}

\vspace*{2mm}
\noindent
{\bf Singlet potential}
\vspace*{2mm}
\\
For the singlet potential, we start from
\be
V_S(r;\mu_1)=- 4 \pi C_F \int_{k>\mu_1} \frac{d^3 k}{(2 \pi)^3} \, e^{i \vec{k} \cdot \vec{r}} \frac{\alpha_{\bz} (k^2)}{k^2} \,  \label{VSmu1}
\ee
in the large-$\bz$ approximation.
Here $C_F=4/3$ is the Casimir operator for the fundamental representation;
$k$ denotes the momentum of the gluon; 
$\alpha_{\bz}$ is an effective coupling appearing after 
the resummation of the perturbative series in the large-$\bz$ approximation,\fn{
We use the modified minimal subtraction ($\overline{\rm MS}$) scheme,
where the one-loop running coupling is obtained as
\be
\alpha_s(\mu)=\frac{4 \pi}{\bz} \frac{1}{\log({\mu^2}/{\LQ^2})} \, , \nonumber
\ee
where $\mu$ is a renormalization scale.
} 
\be
\alpha_{\bz}(k^2)=\frac{4 \pi}{\bz} \frac{1}{\log ({k^2 e^{-5/3}/\LQ^2})} \, ,
\ee
which has a single pole at $k^2=e^{5/3} \LQ^2$;
$\bz=11-\frac{2}{3} n_f$ is the first coefficient of the $\beta$-function with $n_f$ flavors. 
According to Ref.~\cite{Sumino:2005cq},
$V_S(r;\mu_1)$ can be reduced to the form of Eq.~\eqref{VS} with
\be
V_S^{\rm RF}(r)
=\frac{1}{r} \lt[\int_0^{\infty} \frac{d \tau}{\pi \tau} \{-C_F e^{-\sqrt{\tau r^2}}\} {\rm Im} \, \alpha_{\bz}(-\tau+i 0)
-\frac{4 \pi C_F}{\bz} \rt]
+\frac{2 \pi C_F e^{5/3}}{\bz}  \LQ^2 r \, , \label{VSRF}
\ee
and
\be
\mathcal{C}_2(\mu_1)
={\rm Im} \int_{C(\mu_1^2)} \frac{d \tau}{\pi \tau} \lt(-\frac{i}{6} C_F \tau^{3/2} \rt) \alpha_{\bz}(\tau) \, . \label{mu1depcoeff}
\ee
The contour $C(\mu^2)$ is shown in the left panel of Fig.~\ref{fig:contours}.
The cutoff dependence is caused by the $u=3/2$ renormalon,\fn{
Cutoff dependence is generally related to renormalons as explained in Ref.~\cite{Mishima:2016vna}.} 
and shows the sensitivity to the infrared (IR) dynamics.
To eliminate this cutoff dependence, the IR quantity $\delta E_{\US}$ should be added.

\vspace*{2mm}
\noindent
{\bf ${\bm {\delta E_{\US}}}$}
\vspace*{2mm}
\\
The perturbative result of $\delta E_{\US}$ in the large-$\bz$ approximation \cite{Sumino:2004ht}
can be expressed as 
\begin{align}
\delta E_{\US}
=&\frac{C_F}{4 \pi} 8 r^2 \DV^3  [\delta \mathcal{E}^{ren.}_{\US}(\DV)+ \delta \mathcal{E}_{\US}^{\rm LL,div}(\DV;\epsilon) ] \label{EUS} \, ,
\end{align}
where $\delta \mathcal{E}_{\US}^{\rm LL,div}(\DV;\epsilon)$ consists of
the leading log (LL) contribution and ultraviolet (UV) divergences, and 
does not contain renormalons 
[we show the explicit form of $\delta \mathcal{E}_{\US}^{\rm LL,div}(\DV;\epsilon)$ in Eq.~\eqref{EUSLLdiv}].\fn{
The result is given in dimensional regularization where $D=4-2 \epsilon$.}
In contrast, $\delta \mathcal{E}_{\US}^{ren.}$ has renormalons and is given by \cite{Sumino:2004ht}
\be
\delta \mathcal{E}_{\US}^{ren.}=\alpha_s \sum_{n=0}^{\infty} d_n \lt( \frac{\bz \alpha_s}{4 \pi} \rt)^n \label{Etilder} \, ,
\ee
with
\begin{align}
& B(u) \equiv \sum_{n=0}^{\infty} \frac{d_n}{n!} u^n
=\lt(\frac{\DV^2}{\mu^2} e^{-5/3} \rt)^{-u} 
\frac{1}{u} \lt(\frac{1}{2^{2u}}  \frac{2 \Gamma(2-u) \Gamma(2u-3)}{\Gamma(u-1)}-\frac{1}{6} \rt) \, .
\end{align} 

We study the resummation of Eq.~\eqref{Etilder}
starting from its one-dimensional integral representation
with the cutoff scales:
\be
\delta \mathcal{E}_{\US}^{ren.}(\DV; \mu_1,\mu_2)
=\int_{\mu_2^2}^{\mu_1^2} \frac{d \tau}{2 \pi \tau} w_{\US} \lt(\frac{\tau}{\DV^2} \rt) \alpha_{\bz}(\tau) 
\label{start} \, ,
\ee
where $\tau$ is the square of the gluon loop momentum [cf. Fig.~\ref{fig:Diagrams}].
Eq.~\eqref{start} corresponds to integrating out the ultrasoft scale
but keeping the scale $\LQ$ unintegrated.
Here the weight $w_{\US}$ can be calculated as 
\be
w_{\US}(x)=
\begin{cases}
&\frac{\pi}{6} \lt[2+(x-2) \sqrt{1+x} ~\rt] \qquad (x<1)\\
& \frac{\pi}{6} (x-2) \sqrt{1+x} \qquad (x>1)
\end{cases} \, ,
\ee
via the formula in Ref.~\cite{Neubert:1994vb} from the Borel transformation $B(u)$.
The expansion of $w_{\US}(x)$ is given by
\be
w_{\US}(x)
=\begin{cases}
& \frac{\pi}{6} x^{3/2} - \frac{\pi}{4} x^{1/2}+\mathcal{O}(x^{-1/2}) \qquad (x \gg1) \\
& \frac{\pi}{8} x^2+\mathcal{O}(x^3) ~~~~~~~~~~~~~~~~~\qquad (x \ll1)
\end{cases} \, ,
\ee
which agrees with the UV and IR renormalons:
$\mathcal{U}_{\UV}=\lt\{+\frac{3}{2},+ \frac{1}{2}, -\frac{1}{2},\dots \rt\}$, and
$\mathcal{U}_{\IR}=\{2,3,4,\dots \}$.\fn{
The expansion of $w_{\US}(x)$ in $x$ (in particular its power) is related to the IR renormalons and
that in $1/x$ is related to the UV renormalons, as explained in Ref.~\cite{Neubert:1994vb}
in a general context.}
We note that, in the Borel transformation $B(u)$, 
there are positive UV renormalons reflecting 
the strong UV divergences of $\delta E_{\US}$ \cite{Brambilla:1999xf,Sumino:2004ht}.

To extract a renormalon free part, we decompose $\delta \mathcal{E}^{ren.}_{\US}(\DV;\mu_1,\mu_2)$ as follows.
We divide $w_{\US}$ into the UV divergent part and regular part as
\be
w_{\US}(x)=w_{\rm div}(x)+w_{\rm reg}(x) \, ,
\ee
\be
w_{\rm div}(x)=\frac{\pi}{6} x^{3/2} - \frac{\pi}{4} x^{1/2} \, , \label{wdiv}
\ee
where $w_{\rm reg}(x)$ converges to 0 as $x \to \infty$ by construction.
Correspondingly, $\delta \mathcal{E}^{ren.}_{\US}$ is divided as
\be
\delta \mathcal{E}_{\US}^{ren.}(\DV; \mu_1,\mu_2)
=\int_{\mu_2^2}^{\mu_1^2} \frac{d \tau}{2 \pi \tau} w_{\rm div} \lt(\frac{\tau}{\DV^2} \rt) \alpha_{\bz}(\tau)+\int_{\mu_2^2}^{\mu_1^2} \frac{d \tau}{2 \pi \tau} w_{\rm reg} \lt(\frac{\tau}{\DV^2} \rt) \alpha_{\bz}(\tau) \label{divreg} \, .
\ee 
For the first term,
we have
\begin{align}
[\text{1st term in Eq.~\eqref{divreg}}]
&={\rm Im} \, \int_{\mu_2^2}^{\mu_1^2} \frac{d \tau}{2 \pi \tau}  i w_{\rm div}\lt(\frac{\tau}{\DV^2} \rt) \alpha_{\bz}(\tau) \non
&={\rm Im} \lt( \int_{C(\mu_1^2)} - \int_{C(\mu_2^2)} \rt) \frac{d \tau}{2 \pi \tau} i w_{\rm div} \lt(\frac{\tau}{\DV^2} \rt) \alpha_{\bz}(\tau) \, . \label{fromwdiv}
\end{align}

To investigate the second term in Eq.~\eqref{divreg},
we construct a pre-weight $W_{\US}(z)$ satisfying 
$2 \, {\rm Im} W_{\US}(z)=w_{\rm reg}(z)$ for $z \in \mathbb{R}_{\geq 0}$ as \cite{Mishima:2016xuj,Mishima:2016vna}
\be
W_{\US}(z)=\int_0^{\infty} \frac{d x}{2 \pi} \frac{w_{\rm reg}(x)}{x-z-i0} \, ,
\ee
or $W_{\US+}(z) \equiv W_{\US}(-z)$. 
The explicit calculation leads to
\begin{align}
W_{\US+}(z)=
&\frac{\pi}{6} \bigg[\frac{5}{3 \pi}-\frac{3}{4} \sqrt{z}+\frac{z}{\pi}-\frac{1}{2} z \sqrt{z} \non
&+\frac{1}{2 \pi}  (z+2) \sqrt{1-z} \log{\lt( \frac{1-\sqrt{1-z}}{1+\sqrt{1-z}} \rt)}
+\frac{1}{\pi} \log{\lt(\frac{1+z}{z} \rt)} \bigg] .
\end{align}
Using this pre-weight, we can rewrite the second term in Eq.~\eqref{divreg} as
\begin{align}
[\text{2nd term in Eq.~\eqref{divreg}}]
&={\rm Im} \, \int_{\mu_2^2}^{\mu_1^2} \frac{d \tau}{\pi \tau} W_{\US}\lt(\frac{\tau}{\DV^2} \rt) \alpha_{\bz}(\tau) \non
&={\rm Im} \, \lt(\int_{C_{\infty}}-\int_{\mu_1^2}^{\infty}-\int_{C(\mu_2^2)} \rt) 
\frac{d \tau}{\pi \tau} W_{\US}\lt(\frac{\tau}{\DV^2} \rt) \alpha_{\bz}(\tau) \, , \label{regularpart}
\end{align} 
where the contour $C_{\infty}$ is shown in the middle panel of Fig.~\ref{fig:contours}.
In Eq.~\eqref{regularpart}, the integral along $C_{\infty}$ is clearly independent of $\mu_1$ and $\mu_2$.
The second integral is $\mu_1$ dependent.

Although the third integral along $C(\mu^2_2)$ is apparently $\mu_2$ dependent,
we can extract a $\mu_2$-independent part from this integral. 
In evaluating the integral along $C(\mu_2^2)$,
we expand $W_{\US}$ in $\tau/\DV^2 \ll 1$. The expansion is given by
\begin{align}
W_{\US}(z)
&=- \frac{i}{2} w_{\rm div}(z)+\lt(\frac{5}{18}-\frac{\log{2}}{3} \rt)
-\frac{5}{12} z+c_2 z^2+\mathcal{O}(z^{5/2}) \, , \label{expWUS}
\end{align}
where $c_2$ is a polynomial of $\log(z)$ containing a complex number, 
$c_2=-\frac{1}{96} (5+6 \log{(z/4)}-6 i \pi)$.
We note that the integral of the first term ($w_{\rm div}$) 
is canceled against that of Eq.~\eqref{fromwdiv}.
For the $z^0$ and $z^1$-terms with the real coefficients (writing them as $a_n z^n$ for brevity),
we can obtain $\mu_2$-independent results 
\cite{Sumino:2005cq,Mishima:2016xuj,Mishima:2016vna}:
\begin{align}
{\rm Im} \int_{C(\mu_2^2)} \frac{d \tau}{\pi \tau} a_n \lt(\frac{\tau}{\DV^2} \rt)^n \alpha_{\bz}(\tau)
=\frac{1}{2 i} \int_{C_{\LQ}} \frac{d \tau}{\pi \tau} a_n \lt(\frac{\tau}{\DV^2} \rt)^n \alpha_{\bz}(\tau)
=-\frac{4 \pi a_n}{\bz} \lt(\frac{\LQ^2 e^{5/3}}{\DV^2} \rt)^n \label{Cauchy}
\end{align}
since the integrands satisfy $\{f(z)\}^*=f(z^*)$.
The contour $C_{\LQ}$ is shown in the right panel of Fig.~\ref{fig:contours}.
In contrast, the $z^2$-term with the complex coefficient
gives a $\mu_2$-dependent result,
which reduces to $\mathcal{O}(\mu_2^4/\DV^4)$.

Collecting all the contributions to $\delta \mathcal{E}_{\US}^{ren.}$ [Eq.~\eqref{divreg}],  
we reach the following expression:
\be
\delta \mathcal{E}^{ren.}_{\US}(\DV; \mu_1,\mu_2)
=\delta \mathcal{E}^{\rm RF}_{\US}(\DV)
+\delta \mathcal{E}^{\rm UV cut.}_{\US}(\DV;\mu_1)
+\mathcal{O}(\mu_2^4/\DV^4) \, ,
\ee
with
\begin{align}
&\delta \mathcal{E}^{\rm RF}_{\US}(\DV)
={\rm Im} \int_{C_{\infty}} \frac{d \tau}{\pi \tau} W_{\US}\lt(\frac{\tau}{\DV^2} \rt) \alpha_{\bz}(\tau)
+\frac{4 \pi}{\bz} \lt(\frac{5}{18}-\frac{\log{2}}{3} \rt)-\frac{5 \pi}{3 \bz} \frac{e^{5/3} \LQ^2}{\DV^2} \label{RF} \, ,
 \\ 
&\delta \mathcal{E}^{\rm UV cut.}_{\US}(\DV;\mu_1)
={\rm Im} \int_{C(\mu_1^2)}  \frac{d \tau}{2 \pi \tau} i w_{\rm div} \lt(\frac{\tau}{\DV^2} \rt) \alpha_{\bz}(\tau)
-{\rm Im} \int_{\mu_1^2}^{\infty} \frac{d \tau}{\pi \tau} W_{\US}\lt(\frac{\tau}{\DV^2} \rt) \alpha_{\bz}(\tau)  \label{mu1dep} \, .
\end{align}
We have separated the renormalon free part $\delta \mathcal{E}^{\rm RF}_{\US}$
from the UV and IR cutoff dependent parts, 
$\delta \mathcal{E}_{\US}^{\rm UVcut.}$ and $\mathcal{O}(\mu_2^4/\DV^4)$.
With this, the ultrasoft correction $\delta E_{\US}$ is written as [cf. Eq.~\eqref{EUS}] 
\begin{align}
\delta E_{\US}(r;\mu_1,\mu_2)
=\frac{C_F}{4 \pi} 8 r^2 \DV^3 
\bigg[&\delta \mathcal{E}^{\rm RF}_{\US}(\DV)+\delta \mathcal{E}_{\US}^{\rm UVcut.}(\DV;\mu_1) \non
&+\mathcal{O}(\mu_2^4/\DV^4)+\delta \mathcal{E}_{\US}^{\rm LL,div}(\DV;\epsilon) \bigg]  \, . \label{summarize}
\end{align}

One can explicitly see the $u=3/2$ renormalon cancellation
between $\delta E_{\US}$ and the singlet potential $V_S$.
In Eq.~\eqref{mu1dep}, the first term shows a strong $\mu_1$ dependence as 
the integral diverges as $\mu_1 \to \infty$.\fn{
In contrast, the second term has a mild $\mu_1$ dependence.} 
Hence, it is regarded as a UV sensitive part.
In particular, the leading $\mu_1$ dependence of $\delta \mathcal{E}_{\US}^{\rm UVcut.}(\DV;\mu_1)$ gives
\begin{align}
\frac{C_F}{4 \pi} 8 r^2 \DV^3 \delta \mathcal{E}_{\US}^{\rm UVcut.}(\DV;\mu_1)
&
\sim \frac{C_F}{4 \pi} 8 r^2 \DV^3 \, {\rm Im} \int_{C(\mu_1^2)}  \frac{d \tau}{2 \pi \tau} i w_{\rm div} \lt(\frac{\tau}{\DV^2} \rt) \alpha_{\bz}(\tau) \non
& \sim 
\frac{C_F}{4 \pi} 8 r^2 \DV^3 \,
{\rm Im} \int_{C(\mu_1^2)} \frac{d \tau}{2 \pi \tau} i \lt( \frac{\pi}{6} \frac{\tau^{3/2}}{\DV^3} \rt) \alpha_{\bz}(\tau)
=-\mathcal{C}_2(\mu_1) r^2 \, . \label{mu1leadingdep}
\end{align}
See Eq.~\eqref{mu1depcoeff} for $\mathcal{C}_2(\mu_1)$.
As a result, 
this $\mu_1$ dependence cancels 
against the $\mu_1$-dependent $r^2$-term of the singlet potential \eqref{VS} in the multipole expansion \eqref{multi2}.
This is exactly the $u=3/2$ renormalon cancellation.
Once the strong UV divergences of $\delta E_{\US}$ are eliminated in Eq.~\eqref{multi2},\fn{
In fact, the next strongest cutoff dependence caused by the $u=1/2$ renormalon, 
or by the $x^{1/2}$-term of $w_{\rm div}(x)$, remains.
We visit this point in Discussion. \label{fn:u=1/2}} 
we can take the limit $\mu_1 \to \infty$,
where the second term in Eq.~\eqref{mu1dep} vanishes.\fn{
However, $\mu_1$ is originally set as $\mu_1 \ll r^{-1}$,
and the integral vanishes when $\mu_1/\DV \to \infty$.
Hence, the result obtained in this limit corresponds to the case where we consider $r \DV \ll 1$, 
that is $\alpha_s(r^{-1}) \ll 1$.
}
This is similar to 
the situation that the contributions from ordinary negative UV renormalons vanish 
as we take a renormalization scale large.\fn{
The second term of Eq.~\eqref{mu1dep} is identified as the part related to the negative UV renormalons,
since its $\mu_1$ dependence reflects the behavior of $w_{\rm reg}(x)$ at $x \gg 1$,
which is determined by the negative UV renormalons.} 
Therefore,  the $\mu_1$-dependent term $\delta \mathcal{E}_{\US}^{\rm UVcut.}$
does not contribute to the static QCD potential. 

The leading $\mu_2$ dependence of $\delta E_{\US}(r;\mu_1,\mu_2)$
is given by $\mathcal{O}(\mu_2^4 r^2/\DV) \sim \mathcal{O}(\mu_2^4 r^3)$,
which is regarded as a higher order correction.
This IR cutoff dependence is caused by the IR renormalon at $u=2$,
whose connection to the local gluon condensate will be elaborated in Sec.~\ref{sec:LG}.

\begin{figure}[tbhp]
\begin{minipage}{0.5\hsize}
\begin{center}
\includegraphics[width=8cm]{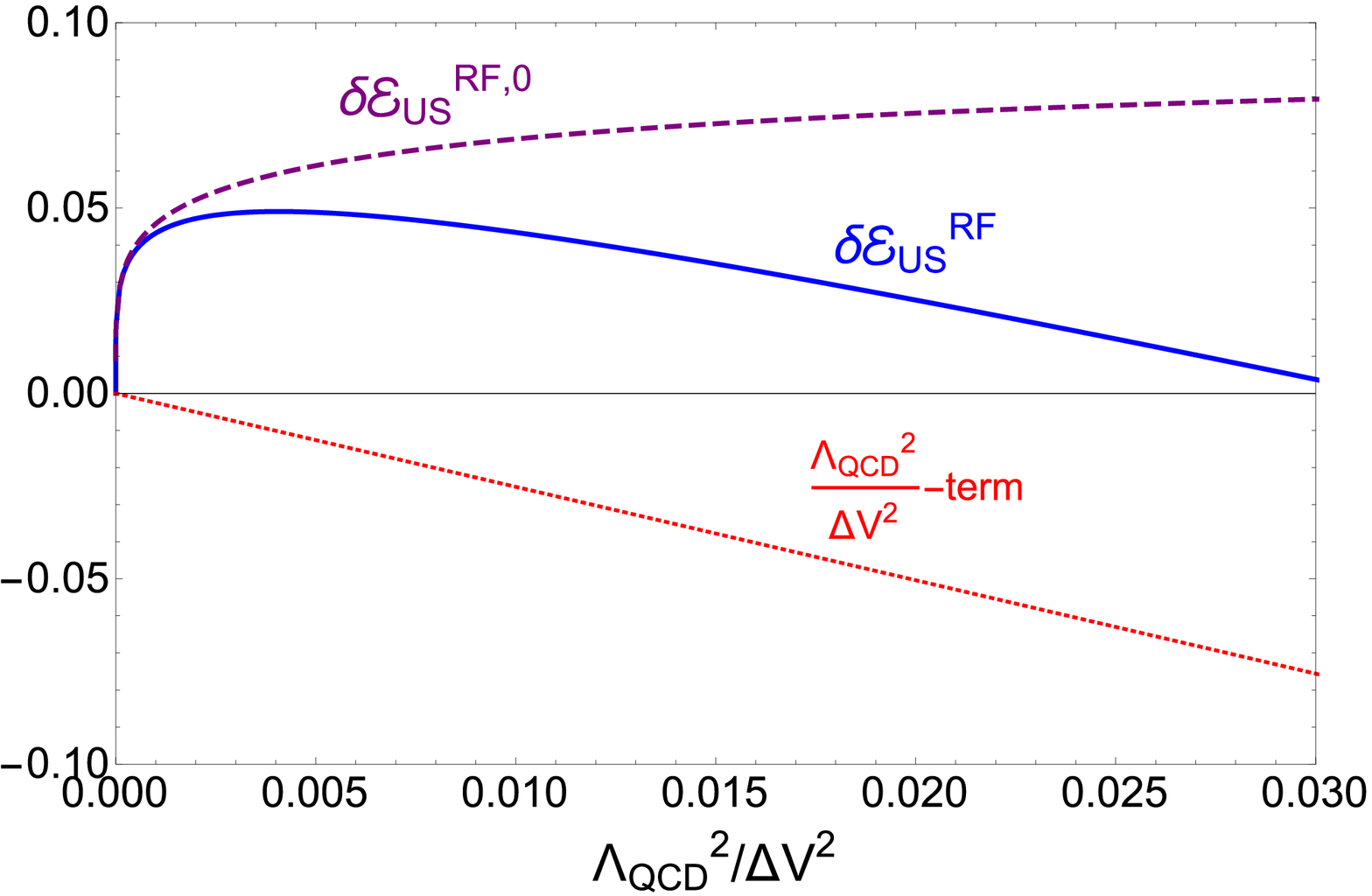}
\end{center}
\end{minipage}
\begin{minipage}{0.5\hsize}
\begin{center}
\includegraphics[width=8cm]{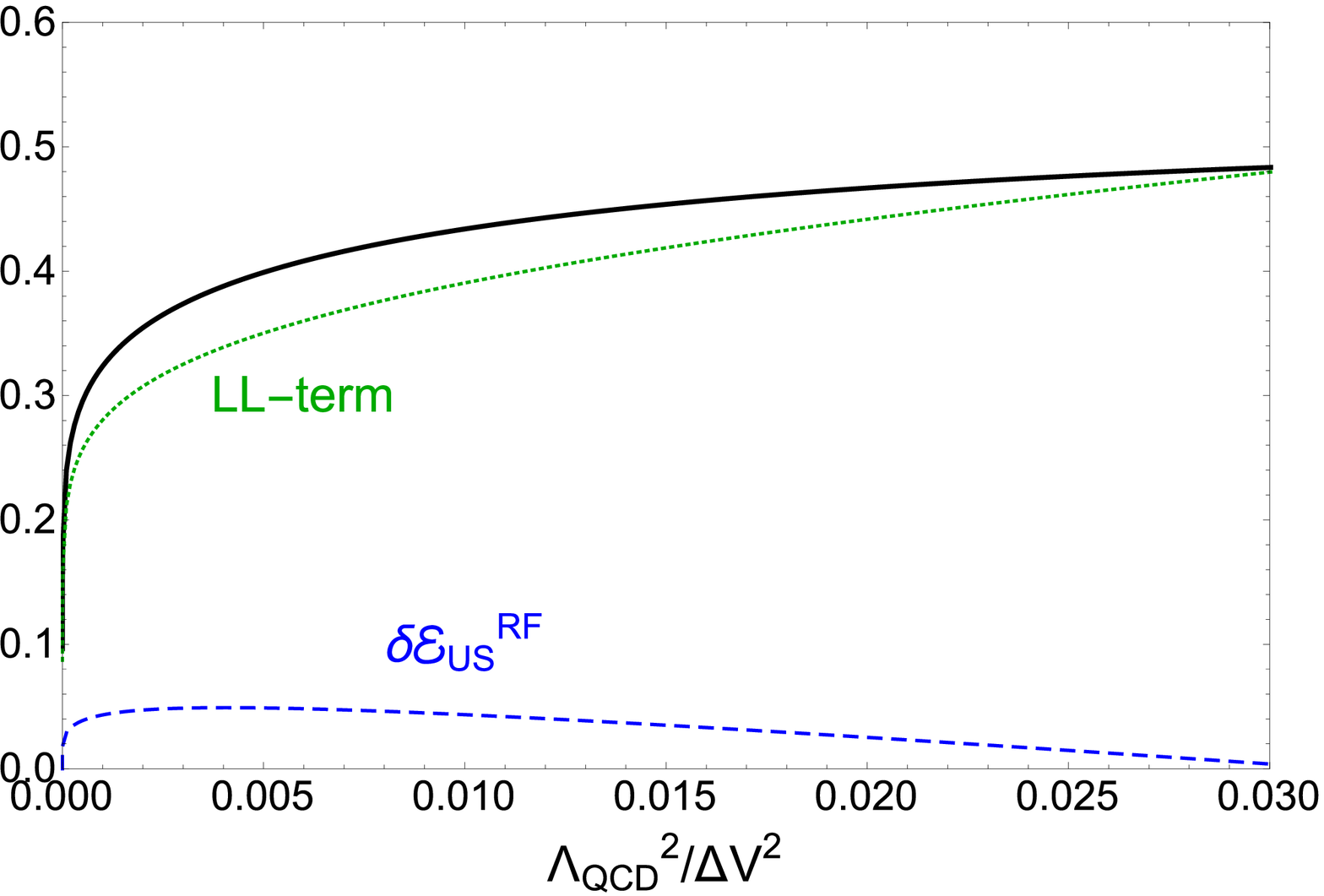}
\end{center}
\end{minipage}
\caption{
(Left) $\delta \mathcal{E}^{\rm RF}_{\US}$ and each term of $\delta \mathcal{E}^{\rm RF}_{\US}$ as functions of $\LQ^2/\DV^2$;
See Eq.~\eqref{RFpart}.
The purple dashed line shows $\delta \mathcal{E}^{\rm RF,0}_{\US}$,
and the red dotted line shows the $\LQ^2/\DV^2$-term.
The sum of them ($\delta \mathcal{E}_{\US}^{\rm RF}$) is represented by the blue solid line.
(Right)
$\delta \mathcal{E}_{\US}^{\rm RF}$ (blue dashed) and 
$\delta \mathcal{E}^{\rm LL,div}_{\US}$ with the divergences omitted (green dotted);
See Eq.~\eqref{EUSLLdiv} and the text below it.
We choose $\alpha_s(\mu)=0.05$. 
The sum of these two parts is shown by the black solid line.  
We set $n_f=0$.
\label{fig:ERFinDV}}
\end{figure}

The renormalon free part of $\delta E_{\US}$, 
$\frac{C_F}{4 \pi} 8 r^2 \DV^3 \delta \mathcal{E}_{\US}^{\rm RF}$, 
is independent of $\mu_1$ and $\mu_2$,
and remains as a net contribution to the static QCD potential.
We rewrite $\delta \mathcal{E}_{\US}^{\rm RF}$ as
\be
\delta \mathcal{E}_{\US}^{\rm RF}(\DV)
=\delta \mathcal{E}_{\US}^{\rm RF,0}(\DV)-\frac{5 \pi}{3 \bz} \frac{e^{5/3} \LQ^2}{\DV^2} \label{RFpart}
\ee
with
\be
\delta \mathcal{E}_{\US}^{\rm RF,0}(\DV)
=\int_0^{\infty} \frac{d \tau}{\pi \tau} W_{\US+} \lt(\frac{\tau}{\DV^2} \rt) {\rm Im} \, \alpha_{\bz}(-\tau+i0)
+\frac{4 \pi}{\bz} \lt(\frac{5}{18}-\frac{\log{2}}{3} \rt) \, ,
\ee
where the path $C_{\infty}$ in Eq.~\eqref{RF} was deformed into the straight line from $\tau=0$ to $-\infty$.
Note that $W_{\US}(z)$ takes a real value for a negative $z$. 
We show the behavior of $\delta \mathcal{E}_{\US}^{\rm RF}$ in Fig.~\ref{fig:ERFinDV}.

\begin{figure}[tbp]
\begin{center}
\includegraphics[width=11cm]{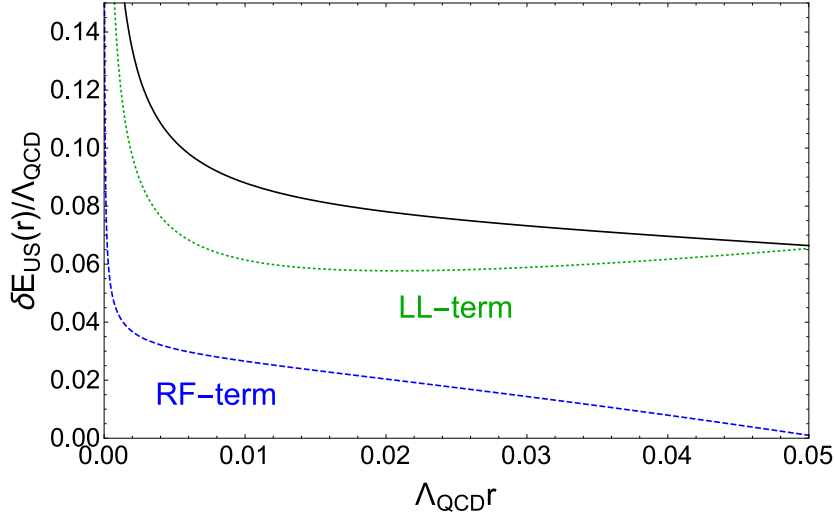}
\end{center}
\caption{
Net $\delta E_{\US}$ correction as a function of $\LQ r$ (black solid).
The contributions from $\delta \mathcal{E}_{\US}^{\rm RF}$ (blue dashed)
and $\delta \mathcal{E}_{\US}^{\rm LL,div}$ (green dotted) are also shown separately.
For $\delta \mathcal{E}_{\US}^{\rm LL,div}$,
we replace $\alpha_s(\mu) \to \alpha_s(r^{-1} e^{-5/6})$ and 
omit the second term in Eq.~\eqref{EUSLLdiv}. 
We set $n_f=0$.  
\label{fig:ERFinR}}
\end{figure}

We include the contribution from $\delta \mathcal{E}^{\rm LL,div}_{\US}$,
which is not taken into account explicitly so far. 
It is given by \cite{Sumino:2004ht}
\begin{align}
\delta \mathcal{E}_{\US}^{\rm LL, div}
&=\alpha_s \sum_{n=0}^{\infty} 
 \lt\{\frac{1}{6} \frac{(-1)^{n+1}}{n+1} \log^{n+1}{ \lt( \frac{\DV^2 e^{-5/3}}{\mu^2} \rt)}
+\frac{(-1)^n}{\epsilon^{n+1}}\frac{g(\epsilon)}{n+1} \rt\} \lt(\frac{\bz \alpha_s}{4 \pi} \rt)^n \non
&=-\frac{2 \pi}{3 \bz} \log{\lt(\frac{\alpha_s(\mu)}{\alpha_s(\DV e^{-5/6})} \rt)} 
+\alpha_s \sum_{n=0}^{\infty} \frac{(-1)^n}{\epsilon^{n+1}}  \frac{g(\epsilon)}{n+1} \lt(\frac{\bz \alpha_s}{4 \pi} \rt)^n \, ,
\label{EUSLLdiv}
\end{align}
where $g(\epsilon)$ is defined in Ref.~\cite{Sumino:2004ht}.
In Fig.~\ref{fig:ERFinDV}, we plot 
$\delta \mathcal{E}_{\US}^{\rm RF}+\delta \mathcal{E}_{\US}^{\rm LL,div}$
by omitting the second term in Eq.~\eqref{EUSLLdiv}
and setting $\alpha_s(\mu)=0.05$.\fn{
As mentioned below, the coupling $\alpha_s(\mu)$ in logarithm is replaced with the coupling at the soft scale
when combined with the singlet potential. 
We choose the value $\alpha_s(\mu)=0.05$ such that
this coupling (supposed to be the coupling at the soft scale) 
is smaller than the coupling at the ultrasoft scale [$\sim \alpha_s(\DV e^{-5/6})$] 
in the range shown in Fig.~\ref{fig:ERFinDV}.
We note that in contrast to the LL terms, the renormalon free part is not affected by the choice of this coupling
 since the resummation of perturbative series has been already performed.
}
Fig.~\ref{fig:ERFinR} shows the 
contribution from $\delta \mathcal{E}_{\US}^{\rm RF}+\delta \mathcal{E}_{\US}^{\rm LL,div}$
to the static QCD potential as a function of $\LQ r$ [cf. Eq.~\eqref{summarize}].
In this figure, 
we replace $\alpha_s(\mu)$ in the first term
of Eq.~\eqref{EUSLLdiv} with $\alpha_s(r^{-1} e^{-5/6})$ \cite{Pineda:2000gza}
and omit the second term.\fn{
This prescription is equivalent to assuming that 
the counter term is provided as
\be
-\frac{C_F}{4 \pi} 8 r^2 \DV^3 \lt[
\alpha_s \sum_{n=0}^{\infty} 
\frac{(-1)^n}{\epsilon^{n+1}}  \frac{g(\epsilon)}{n+1} \lt(\frac{\bz \alpha_s}{4 \pi} \rt)^n 
-\frac{2 \pi}{3 \bz} \log{\lt(\frac{\alpha_s(\mu)}{\alpha_s(r^{-1} e^{-5/6})} \rt)} \rt] \, ,
\ee
from the singlet potential. At LO, this counter term gives  
$-\frac{C_F}{4 \pi} 8 r^2 \DV^3 \alpha_s \lt[\frac{1}{6 \epsilon}+\frac{1}{3} \log(\mu r) \rt] $.
In a natural scheme where one minimally subtracts the divergence and the associated logarithm 
of the soft contribution at 3-loop in momentum space 
(as adopted in Ref.~\cite{Anzai:2010td}), the counter term is given by
$
-\frac{C_F}{4 \pi} 8 r^2 \DV^3 \alpha_s \lt[\frac{1}{6 \epsilon}+\frac{1}{3} \log(\mu r)+\frac{\gamma_E}{3} \rt] \, .
$ 
\label{fn:finite}} 
In evaluating the $r$-dependence of $\DV$, 
we substitute the LL result $\DV=\frac{C_A \alpha_s(r^{-1})}{2 r}$ with $C_A=3$.\fn{
We notice that the three-loop result for $\DV$ is currently available \cite{Anzai:2009tm,Smirnov:2009fh,Anzai:2013tja,Lee:2016cgz},
while we used the LL result for simplicity.}
One can see in Fig.~\ref{fig:ERFinR} that the ultrasoft correction 
shows a decreasing behavior, 
while the singlet potential exhibits the opposite behavior.

\begin{figure}[tbp]
\begin{center}
\includegraphics[width=11cm]{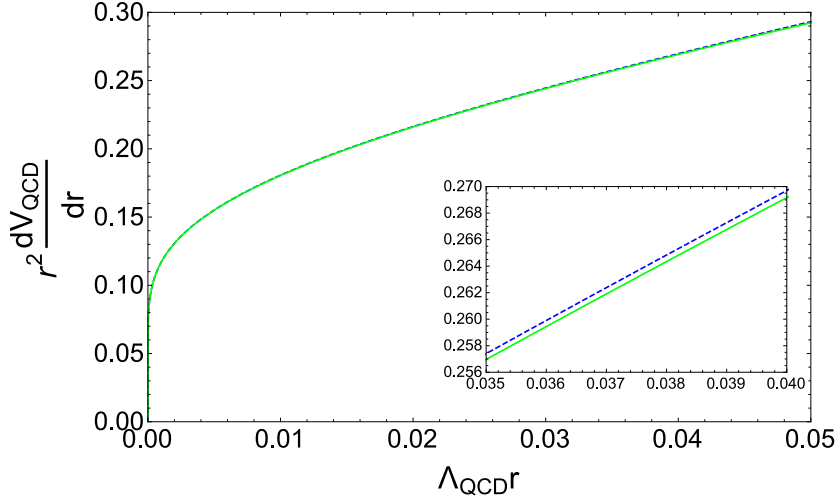}
\end{center}
\caption{
Dimensionless forces (with a minus sign)
given by $V_S(r)^{\rm RF}$ [Eq.~\eqref{VSRF}] (blue dashed) 
and given by the sum of $V_S^{\rm RF}$ and the net $\delta E_{\US}$ contribution (green solid).
We set $n_f=0$.
\label{fig:Force}}
\end{figure}

We also examine how the net ultrasoft correction 
modifies the result of the soft contribution \eqref{VSRF}.
In Fig.~\ref{fig:Force}, we plot the dimensionless QCD force (with a minus sign) 
before and after adding the ultrasoft correction from
$\delta \mathcal{E}_{\US}^{\rm RF}+\delta \mathcal{E}_{\US}^{\rm LL,div}$.
The prescription for $\delta \mathcal{E}_{\US}^{\rm LL,div}$
is the same as in Fig.~\ref{fig:ERFinR}.
In Fig.~\ref{fig:Force}, the ultrasoft correction is negative and
its relative size is 0.1 \% level. 

\section{Renormalon free definition of local gluon condensate}
\label{sec:LG}
The separation of $\delta E_{\US}$ enables us to define the local gluon condensate in a renormalon free way.
We are considering the hierarchy $\LQ \ll \alpha_s(r^{-1})/r \ll r^{-1}$ 
and already integrated out the mode $k \sim \alpha_s(r^{-1})/r$ in $\delta E_{\US}$.
Hence, the remaining mode contributing to $\delta E_{\US}$ is specified by the $\LQ$ scale.
In this case, $\delta E_{\US}$ of the form of Eq.~\eqref{deltaEUS} turns to the local gluon condensate
due to the hierarchy $\Delta V \gg k \sim \LQ$, where $k$ denotes the energy scale of the active gluon \cite{Flory:1982qx,Voloshin:1978hc,Brambilla:1999xf,Sumino:2014qpa}:
\be
\delta E_{\US}(k<\mu_2)=\frac{T_F}{12 N_c} \frac{r^2}{\DV} \braket{0|g^2 G^{a \mu \nu} G^a_{\mu \nu}|0}|_{k < \mu_2}
+\mathcal{O}(1/\DV^2) \, .
\ee
Including this effect, the static QCD potential is given by 
\be
V_{\rm QCD}(r)
=V_S(r;\mu_1)+\delta E_{\US}(r;\mu_1,\mu_2)+\frac{T_F}{12 N_c} \frac{r^2}{\DV} \braket{0|g^2 G^{a \mu \nu} G^a_{\mu \nu}|0}|_{k < \mu_2} \, . \label{Threeterms}
\ee
The $\mu_2$ dependence of the nonperturbative term [the last term in Eq.~\eqref{Threeterms}] 
is evaluated as
\be
\frac{T_F}{12 N_c} \frac{r^2}{\DV} \braket{0|g^2 G^{a \mu \nu} G^a_{\mu \nu}|0}|_{k < \mu_2}
\sim \frac{C_F}{8 \pi} \frac{r^2}{\DV} \int^{\mu_2^2} d \tau \, \tau \alpha_{\bz}(\tau)
\ee
based on perturbation theory in the large-$\bz$ approximation.\fn{
For the region $k \sim \mu_2$, perturbation theory is expected to still work since $\mu_2 \gg \LQ$.}
In regularizing the local gluon condensate, 
we used a naive point splitting.

The leading $\mu_2$ dependence in $\delta E_{\US}(r;\mu_1,\mu_2)$
provides a counterpart of the above $\mu_2$ dependence.
The leading $\mu_2$ dependence is explicitly given by 
\be
\delta E_{\US}(r;\mu_1,\mu_2) 
\sim -\frac{C_F}{8 \pi} \frac{r^2}{\DV} \int^{\mu_2^2} d \tau \, \tau \alpha_{\bz}(\tau) \, , \label{EUSIRcut}
\ee
which stems from the integral along $C(\mu_2^2)$ in Eq.~\eqref{regularpart} 
with the $c_2 z^2$-term in Eq.~\eqref{expWUS} considered. 
Therefore, by including the leading $\mu_2$ dependence of $\delta E_{\US}(r;\mu_1,\mu_2)$
in the third term of Eq.~\eqref{Threeterms},
the local gluon condensate can be defined in a factorization scale ($\mu_2$) independent way.
This quantity corresponds to a net local gluon condensate free from the $u=2$ renormalon.
It does not have the instability due to the artificial factorization scale any more
and would depend purely on $\LQ$.

In this way, we can obtain the expression
where the local gluon condensate is contained in a renormalon free way. 
By comparing this result with, for instance, the lattice data for the static QCD potential,
one can extract the value of the local gluon condensate
without being annoyed from the $u=2$ renormalon.

\section{Discussion}
In this section, we discuss the divergences which are not removed in this work
and examine the validity of the large-$\bz$ approximation.

It is known that 
the singlet potential has IR divergences
from the three-loop order in perturbation theory \cite{Appelquist:1977tw,Brambilla:1999qa}.
However, in our calculation, they do not appear due to the large-$\bz$ approximation.
On the other hand, 
$\delta E_{\US}$ in the large-$\bz$ approximation
contains the UV divergences as in Eq.~\eqref{EUSLLdiv}, 
which are the counterpart of the IR divergences of the singlet potential. 
Due to this mismatch in the large-$\bz$ approximation,
only the UV divergences of the ultrasoft correction are left. 
To cancel these UV divergences, 
we should include the IR divergent contributions
to the singlet potential as discovered in Ref.~\cite{Appelquist:1977tw}.
Once they are included, 
the divergences vanish and
$\alpha_s(\mu)$ in logarithm in Eq.~\eqref{EUSLLdiv}
should be replaced with $\alpha_s(r^{-1} e^{-5/6})$ 
as studied in Ref.~\cite{Pineda:2000gza}.
We drew Figs.~\ref{fig:ERFinR} and \ref{fig:Force} based on this expectation.
For a finite part which may remain after this cancellation, see footnote \ref{fn:finite}.

Although the $u=3/2$ renormalon of $\delta E_{\US}$ gets correctly canceled as we observed, 
the $u=1/2$ renormalon of $\delta E_{\US}$ remains uncanceled.
Namely, the cutoff dependence $\sim \mu_1 r^2 \DV^2$ is left,\fn{
This suggests that the $u=1/2$ renormalon of $\delta E_{\US}$ does not correspond to that of $V_S$,
which gives an $r$-independent dependence $\sim \mu_1$.} 
which is caused by the $x^{1/2}$-term of $w_{\rm div}(x)$ in the first line of Eq.~\eqref{mu1leadingdep}.
This dependence is subleading compared to that of the $u=3/2$ renormalon [$\mathcal{O}(\mu_1^3 r^2)$]
due to $\DV/\mu_1 \ll 1$.
However, it still has a positive power dependence on $\mu_1$, and hence, should be removed.
While this renormalon cancellation has not been confirmed,\fn{
The $u=1/2$ renormalon in $\delta E_{\US}$ would be canceled against
the perturbative series whose leading contribution originates from the 2-loop diagrams 
for the singlet potential where three gluon lines appear.
This can be seen from the power of $\alpha_s$ and the color factors.
However, the confirmation of this renormalon cancellation using the large-$\bz$ approximation is not straightforward
since this approximation is usually applied for single gluon exchanging diagrams.}  
it is expected to get canceled against the singlet potential 
when one goes beyond the large-$\bz$ approximation.
This is because the QCD potential is originally $\mu_1$ independent.
Although our argument proceeded while assuming that the $u=1/2$ renormalon vanishes in the end,
we might have a finite contribution after this renormalon cancellation,
which is not included in this Letter.

Before closing this section, we check the validity of the large-$\bz$ approximation for $\delta E_{\US}$.
$\delta E_{\US}$ has been calculated up to NLO in perturbation theory in Ref.~\cite{Pineda:2011aw}.
Apart from the poles in $1/\epsilon$, we have \cite{Brambilla:2006wp, Pineda:2011aw}
\begin{align}
\delta E_{\US}|_{\rm NLO}^{\rm exact}
=C_F \DV^3 r^2 &
\bigg[\{0.030-0.212 \log(\DV/\mu)\} \alpha_s \non
&+\{0.473-1.248 \log(\DV/\mu)+0.186 \log^2(\DV/\mu) \} \alpha_s^2 \bigg] \, ,
\end{align}
while in the large-$\bz$ approximation we have
\begin{align}
\delta E_{\US}|^{\rm large-\bz}_{\rm NLO}
=C_F \DV^3 r^2 &
\bigg[\{0.030-0.212 \log(\DV/\mu)\} \alpha_s \non
&+\{0.657-0.362 \log(\DV/\mu)+0.186 \log^2(\DV/\mu) \} \alpha_s^2 \bigg] \, ,
\end{align}
for $n_f=0$.
For instance, for $\mu=\DV$ and $\alpha_s=0.1$, we have 
$\delta E_{\US}|_{\rm NLO}^{\rm large-\bz}/\delta E_{\US}|_{\rm NLO}^{\rm exact}=1.239$.

\section{Conclusions}
In the multipole expansion of the static QCD potential,
we separated the NLO term, $\delta E_{\US}$, 
into renormalon uncertainties and a renormalon free part.
We focused on the very short distances and
used the large-$\bz$ approximation in the perturbative evaluation of $\delta E_{\US}$.
Owing to the separation, we observed the $u=3/2$ renormalon cancellation between
the soft quantity $V_S$ and the ultrasoft quantity $\delta E_{\US}$ in an explicit way. 
The NLO result ($V_S+\delta E_{\US}$) was presented 
by the sum of the renormalon free parts of $V_S$ and $\delta E_{\US}$. 

After the $u=3/2$ renormalon is removed, 
the leading uncertainty of the NLO calculation related to the IR structure 
is caused by the $u=2$ renormalon.
This is compatible with the fact that the first nonperturbative effect is given by the local gluon condensate.
We explicitly confirmed within the large-$\bz$ approximation
that the $u=2$ renormalon of $\delta E_{\US}$ cancels against that of the local gluon condensate. 
As a result, we obtained the expression of the $r$ expansion where
the local gluon condensate is included in a renormalon free way.
Such a result should be useful to extract a value of the local gluon condensate
numerically without suffering from the $u=2$ renormalon uncertainty.  
We remark that, in order to determine the local gluon condensate with high accuracy,
the formulation presented here requires to be further developed beyond the large-$\bz$ approximation.

\section*{Acknowledgements}

The author is very grateful to Yukinari Sumino for 
private communication on related topics and giving useful comments on the manuscript.

\bibliographystyle{utphys}
\bibliography{BibPhD}
\end{document}